\documentstyle[amsfonts,preprint,aps]{revtex}
\tightenlines
\begin{document}
\title{Theory of diffraction for 2D photonic crystals with a boundary }
\author{D. Felbacq, E. Centeno}
\address{LASMEA UMR-CNRS 6602\\
Complexe des C\'{e}zeaux\\
63177 Aubi\`{e}re Cedex\\
France}
\maketitle

\begin{abstract}
We extend a modal theory of diffraction by a set of parallel fibers to deal
with the case of a hard boundary: that is a structure made for instance of
air-holes inside a dielectric matrix. Numerical examples are given
concerning some resonant phenomena.
\end{abstract}

\section{Introduction}

Photonic crystals (PCs) are periodically modulated structures that present
the properties of having photonic band gaps \cite{joan,weisbuch,weis}. In
the case of 2D photonic crystal where it is possible to study separately $s$%
- and $p$-polarized fields, it is now known that large gaps are easier to
obtain with inverted-contrast crystals (i.e. air holes in a dielectric
matrix) with $p$-polarized fields than with $s$-polarized fields \cite
{cent1,mays1}. In this work we extend a multi-scattering theory by cylinders 
\cite{cent1,tayeb,moi,niak,sabouroux} in order to deal with
inverted-contrast crystals with a hard boundary, that is a structure made of
dielectric or metallic inclusions in a dielectric matrix embedded in vacuum
(fig. 1). This theory allows to quantify the importance of the hard boundary
when the device is considered as a 2D PC and also to study the propagation
phenomena when the device modelizes a photonic crystal fiber \cite
{fibre,fibre2,fibre3,fibre4}.

\section{Theory of diffraction}

We start by constructing the generalized scattering matrix of a set of $N$
dielectric or metallic rods embedded in a dielectric cylinder of circular
cross section $\Omega $ (fig.1) and radius $R$. We use a Cartesian
coordinate system $(O;x,y,z)$ with origin at the center of $\Omega $. The
electromagnetic fields considered here are harmonic fields with a time
dependence of $\exp (-i\omega t)$.

The rods are denoted by $\left\{ D_{n}\right\} $, they are filled with a
dielectric (relative permittivity $\varepsilon _{n}$) or metallic material.
The cylinder in which the fibers are contained has relative permittivity $%
\varepsilon _{r}$, finally this cylinder is embedded in vacuum. Due to the
invariance of the medium along the $z$-direction we look for solutions of
Maxwell equations with a $z$ dependence of $\exp \left( i\gamma z\right) $.
Under this assumption, it is easily shown that all the components of both
magnetic and electric fields are known once $E_{z}$ and $H_{z}$ are known.
Denoting

\[
{\bf F}=\left( 
\begin{array}{c}
E_{z} \\ 
H_{z}
\end{array}
\right) , 
\]
the following propagation equation is satisfied 
\[
\Delta _{\bot }{\bf F+}\chi ^{2}{\bf F=}0 
\]
we denote $\Delta _{\bot }=\frac{\partial ^{2}}{\partial x^{2}}+\frac{%
\partial ^{2}}{\partial y^{2}}$, $\chi ^{2}=k^{2}\varepsilon \left( x\right)
-\gamma ^{2}$ and $k$ is the wavenumber. We denote 
\[
\left\{ 
\begin{array}{l}
\chi _{r}^{2}=k^{2}\varepsilon _{r}-\gamma ^{2} \\ 
\chi _{n}^{2}=k^{2}\varepsilon _{n}-\gamma ^{2} \\ 
\chi _{0}^{2}=k^{2}\varepsilon _{0}-\gamma ^{2}
\end{array}
\right. 
\]
The total exterior field ${\bf F}$ is expanded in Fourier-Bessel expansion
outside $\Omega $%
\begin{equation}
{\bf F}\left( r,\varphi \right) =\sum_{m}\left[ {\bf F}_{m}^{i,+}J_{m}\left(
\chi _{0}r\right) +{\bf F}_{m}^{d,+}H_{m}^{\left( 1\right) }\left( \chi
_{0}r\right) \right] e^{im\varphi },\text{ }\left| r\right| \geq R
\label{out}
\end{equation}
{\bf Remark}: $\sum_{m}{\bf F}_{m}^{i,+}J_{m}\left( \chi _{0}r\right)
e^{im\varphi }$ represents the incident field.

The total field ${\bf F}$ inside $\Omega $ writes in the vicinity of the
boundary $\partial \Omega $ of $\Omega $ 
\begin{equation}
{\bf F}\left( r,\varphi \right) =\sum_{m}\left[ {\bf F}_{m}^{i,-}J_{m}\left(
\chi _{r}r\right) +{\bf F}_{m}^{d,-}H_{m}^{\left( 1\right) }\left( \chi
_{r}r\right) \right] e^{im\varphi }  \label{in}
\end{equation}
We denote 
\[
\widehat{{\bf F}}^{i,\pm }=\left( {\bf F}_{m}^{i,\pm }\right) _{m\in {\Bbb Z}%
},\text{ }\widehat{{\bf F}}^{d,\pm }=\left( {\bf F}_{m}^{d,\pm }\right)
_{m\in {\Bbb Z}} 
\]
so that the transmission conditions on $\partial \Omega $ write 
\begin{equation}
\begin{array}{rrr}
\widehat{{\bf F}}^{i,+}+\widehat{{\bf F}}^{d,+} & = & \widehat{{\bf F}}%
^{i,-}+\widehat{{\bf F}}^{d,-} \\ 
{\bf L}_{\varphi }^{+}\left[ \widehat{{\bf F}}^{i,+}+\widehat{{\bf F}}^{d,+}%
\right] & = & {\bf L}_{\varphi }^{-}\left[ \widehat{{\bf F}}^{d,+}+\widehat{%
{\bf F}}^{d,-}\right]
\end{array}
\label{raccord}
\end{equation}
where ${\bf L}_{\varphi }^{\pm }$ are boundary impedance operators easily
deduced from Maxwell equations. Conditions (\ref{raccord}) lead to 
\begin{equation}
\left( 
\begin{array}{c}
\widehat{{\bf F}}^{i,-} \\ 
\widehat{{\bf F}}^{d,+}
\end{array}
\right) =\left[ 
\begin{array}{cc}
{\cal S}_{1}^{-} & {\cal S}_{2}^{-} \\ 
{\cal S}_{1}^{+} & {\cal S}_{2}^{+}
\end{array}
\right] \left( 
\begin{array}{c}
\widehat{{\bf F}}^{i,+} \\ 
\widehat{{\bf F}}^{d,-}
\end{array}
\right)  \label{system}
\end{equation}
where ${\cal S}_{1,2}^{\pm }$ are linear operators deduced from (\ref
{raccord}) linking the incoming and outgoing parts of the fields on the
boundary of $\Omega $.

{\bf Remark}: In case where the boundary is not circular the theory can
still be applied provided the expansions (\ref{out}-\ref{in}) are restricted
respectively to the smallest circle of center $O$ containing $\Omega $ and
to the greatest circle contained in $\Omega $. In that case operators ${\cal %
S}_{1,2}^{\pm }$ have to be computed numerically (for instance using the
Method of Fictitious Sources \cite{zolla} or the Differential Method \cite
{vincent}).

Around each rod $D_{n}$ the diffracted part of the field has the following
expansion: $\sum\limits_{k}{\bf F}_{n,k}^{d,-}H_{k}^{\left( 1\right) }\left(
\chi r_{n}\right) e^{ik\varphi _{n}}$ where $\left( r_{n},\varphi
_{n}\right) $ are the polar coordinates associated to fiber $n$. The Hankel
part of the total field is the sum of the fields diffracted by each rod $%
D_{n}$ inside $\Omega $: 
\begin{equation}
\sum_{m}{\bf F}_{m}^{d,-}H_{m}^{\left( 1\right) }\left( \chi r\right)
e^{im\varphi }=\sum_{n=1}^{N}\sum_{l}{\bf F}_{n,l}^{d}H_{l}^{\left( 1\right)
}\left( \chi r_{n}\right) e^{il\varphi _{n}}
\end{equation}
so that there is a linear operator ${\cal L=}\left( {\cal L}_{m}^{n}\right)
_{m\in {\Bbb Z}}^{n=1..N}$ such that ${\bf F}_{m}^{d,-}=\sum_{n=1}^{N}{\cal L%
}_{m}^{n}\left[ \widehat{{\bf F}}_{n}^{d}\right] $ where $\widehat{{\bf F}}%
_{n}^{d}=\left( {\bf F}_{n,l}^{d}\right) _{l\in {\Bbb Z}}$

Conversely, there is a linear operator ${\cal R=}\left( {\cal R}%
_{n}^{m}\right) _{n=1..N}^{m\in {\Bbb Z}}$ , obtained from the translation
formula for Bessel functions \cite{abra},\ such that ${\cal R}_{n}^{l}\left( 
\widehat{{\bf F}}^{i,-}\right) ={\bf F}_{l,n}^{i,-}$. Operator ${\cal R}$ is
injective, i.e. it is left invertible, whereas ${\cal L}$ is surjective,
i.e. it is right invertible: ${\cal L\times R}=N$ ${\Bbb I}_{d}.$ The
multi-scattering theory without boundary \cite
{cent1,tayeb,moi,niak,sabouroux} shows that it\ is possible to define a
linear operator ${\cal H}$ such that 
\begin{equation}
{\cal H}^{-1}\left( 
\begin{array}{c}
\widehat{{\bf F}}_{1}^{d} \\ 
\vdots \\ 
\widehat{{\bf F}}_{N}^{d}
\end{array}
\right) ={\cal R}\widehat{{\bf F}}^{i,-}
\end{equation}
from which we derive the scattering matrix of the system of rods in the
absence of a boundary, that is when the medium of permittivity $\varepsilon
_{r}$ fills the entire space: ${\cal S}_{wb}{\cal =LHR}$, which satisfies $%
\widehat{{\bf F}}^{d,-}={\cal S}_{wb}\widehat{{\bf F}}^{i,-}$. System (\ref
{system}) then rewrites 
\begin{equation}
\left( 
\begin{array}{c}
\widehat{{\bf F}}^{i,-} \\ 
\widehat{{\bf F}}^{d,+}
\end{array}
\right) =\left[ 
\begin{array}{cc}
{\cal S}_{1}^{-} & {\cal S}_{2}^{-}{\cal S}_{wb} \\ 
{\cal S}_{1}^{+} & {\cal S}_{2}^{+}{\cal S}_{wb}
\end{array}
\right] \left( 
\begin{array}{c}
\widehat{{\bf F}}^{i,+} \\ 
\widehat{{\bf F}}^{i,-}
\end{array}
\right)
\end{equation}
from which we derive the expression of the internal and exterior fields from
the incident field: 
\begin{equation}
\left\{ 
\begin{array}{l}
\widehat{{\bf F}}^{i,-}=\left( I_{d}+{\cal S}_{2}^{-}{\cal S}_{wb}\right)
^{-1}{\cal S}_{1}^{-}\widehat{{\bf F}}^{i,+} \\ 
\widehat{{\bf F}}^{d,-}={\cal S}_{wb}\left( I_{d}+{\cal S}_{2}^{-}{\cal S}%
_{wb}\right) ^{-1}{\cal S}_{1}^{-}\widehat{{\bf F}}^{i,+} \\ 
\widehat{{\bf F}}^{d,+}=\left[ {\cal S}_{1}^{+}+{\cal S}_{2}^{+}{\cal S}%
_{wb}\left( I_{d}+{\cal S}_{2}^{-}{\cal S}_{wb}\right) ^{-1}{\cal S}_{1}^{-}%
\right] \widehat{{\bf F}}^{i,+}
\end{array}
\right.
\end{equation}
and the problem is solved. This formulation allows to use distinct numbers $%
n_{c}$ (exterior cylinder) and $n_{f}$ (fibers) of Fourier-Bessel
coefficients. This point is a crucial advantage for a low computation time.
Indeed, for a given wavelength the interior diffraction problem may be
correctly described with a small number of Fourier-Bessel coefficients $%
n_{f} $ whereas the exterior problem requires a larger number of coefficient 
$n_{c} $.

In the case of a diffraction problem, $\widehat{{\bf F}}^{i,+}$ represents
the incident field illuminating the structure such as a plane wave, a
gaussian beam or a cylindrical wave. The plane wave defined by the trihedron 
$({\bf E}^{i},{\bf H}^{i},{\bf k}_{0})$ is spatially characterized by its
Euler angles $\left( \varphi _{0},\theta _{0},\delta _{0}\right) $ where $%
\varphi _{0},$ $\theta _{0}$ and $\delta _{0}$ are respectively called
precession, conicity and polarization angle, see fig. 2. Therefore, the
Cartesian coordinates of ${\bf k}_{0}$ are given by:

\[
{\bf k}_{0}{\bf =}\left\{ 
\begin{array}{l}
k_{0}\sin \theta _{0}\cos \varphi _{0} \\ 
k_{0}\sin \theta _{0}\sin \varphi _{0} \\ 
k_{0}\cos \theta _{0}
\end{array}
\right. 
\]
In this formalism, the case of $s$ (resp. $p$) polarization is defined by
the parameters $\theta _{0}=90%
%TCIMACRO{\unit{\UNICODE{0xb0}}}%
%BeginExpansion
\mathop{\rm %
{{}^\circ}}%
%EndExpansion
$ and $\delta _{0}=90%
%TCIMACRO{\unit{\UNICODE{0xb0}}}%
%BeginExpansion
\mathop{\rm %
{{}^\circ}}%
%EndExpansion
$ (resp. $\delta _{0}=0%
%TCIMACRO{\unit{\UNICODE{0xb0}}}%
%BeginExpansion
\mathop{\rm %
{{}^\circ}}%
%EndExpansion
$). This corresponds to the cases where the unique non zero component of the
electric field (resp magnetic field) is $E_{z}$ (resp. $H_{z}$).

\section{Numerical example}

We consider a hexagonal photonic crystal constituted by $19$ air holes
embedded in a dielectric circular cylinder of optical index $\sqrt{%
\varepsilon _{r}}=4$, see fig. 3. The radii of the holes and of the cylinder
core are respectively equal to $r_{f}=0.8$ and $R=10$. The two-dimensional
structure is illuminated by a plane wave in $p$-polarization. The
transmission coefficient $T$ is defined as the flux of the Poynting vector
of the total field collected on a segment situated below the cylinder sheath
and normalized to the incident energy, see fig 3. The convergence of the
vector multi-scattering method is studied with respect to the numbers $n_{f}$
and $n_{c}$ of Fourier coefficients respectively used for the series
expansion of the interior (fibers) and exterior (cylinder sheath) problems.
Figure 4 presents the relative error of the transmission $T$ versus $n_{f}$
and $n_{c}$ for an incident plane wave with $\lambda /R=1.70$ i.e. in the
resonant domain. The vector multi-scattering method converges with a
relative error less than $0.1\%$ with $n_{c}>15$ and for $n_{f}$ higher than 
$3$. However, independently of $n_{c}$, the number of Fourier coefficients $%
n_{h}$ must be higher than $4$ for a correct convergence of the diffraction
problem. Therefore, in order to get a good numerical convergence in this
wavelength domain, we choose $n_{c}=20$ and $n_{f}=4$. In that case, using a
Personnal Computer with a 200 MHz processor and with 64 Mo of RAM, the
computation time for the transmission coefficient for a wavelength and for
the structure defined in fig.3 is about $50$ seconds. We have also tested
our results against that obtained using the fictitious Sources Method \cite
{zolla,zolla2}.

We now study the scattering properties of the PC embedded inside the
cylinder sheath defined in fig.3:\newline
We start with a hexagonal PC constituted by $19$ air holes embedded in an
infinitely dielectric medium of optical index $\sqrt{\varepsilon _{r}}=4$
(i.e. without the circular boundary). The PC presents a photonic band gap
for the interval of wavelengths $\lambda /R=[2.0;2.70]$ for $p$%
-polarization, see the bold curve of figure 5. The solid curve represents
the transmission $T$ of the PC without boundary and doped by a central
microcavity (the central air hole is removed) computed for $p$-polarization (%
$\theta _{0}=90%
%TCIMACRO{\UNICODE[m]{0xb0}}%
%BeginExpansion
{{}^\circ}%
%EndExpansion
$ and $\delta _{0}=0%
%TCIMACRO{\UNICODE[m]{0xb0}}%
%BeginExpansion
{{}^\circ}%
%EndExpansion
$). It appears a resonant wavelength $\lambda _{r}(90%
%TCIMACRO{\UNICODE[m]{0xb0}}%
%BeginExpansion
{{}^\circ}%
%EndExpansion
)/R=2.13$ inside the photonic band gap associated with a localized mode. 
\newline
Let us now compare these results with that corresponding to the structure
defined in fig.3 (PC with the circular boundary):\newline
Figure 6 gives the transmission diagram versus the wavelength for a $p$%
-polarized incident plane wave. The solid curve represents the transmission
in the case where the central air hole is removed whereas the dashed curve
represents the transmission for the perfect crystal (with the central hole).
This structure also presents a photonic band gap for the interval $%
[2.15;2.7] $ but slightly shifted toward higher wavelengths in comparison
with the PC without the circular boundary. We can also remark that though
the cylinder sheath is illuminated in the resonant domain, the photonic band
gap phenomenon provokes the extinction of\ the electromagnetic modes of the
exterior cylinder.\newline
The doped PC embedded inside the cylinder sheath presents a maximum of
transmission inside the photonic band gap domain for the wavelength $\lambda
_{r}^{c}(90%
%TCIMACRO{\UNICODE[m]{0xb0}}%
%BeginExpansion
{{}^\circ}%
%EndExpansion
)/R=2.20$. The map of the modulus of the magnetic field confirms that the
resonant wavelength $\lambda _{r}^{c}(90%
%TCIMACRO{\UNICODE[m]{0xb0}}%
%BeginExpansion
{{}^\circ}%
%EndExpansion
)/R=2.20$ is associated to a localized mode of the structure, see fig.7. In
conclusion, {\bf the resonant wavelengths of PCs inside an infinite
dielectric medium are shifted toward higher wavelengths when PCs are
embedded inside a cylinder sheath.} These results demonstrate that rigorous
numerical computations of finite-size structures are necessary to
efficiently describe diffraction properties of PC fibers.

Recent studies have demonstrated that resonant wavelengths associated to
localized modes strongly depend upon the propagation coefficient $\gamma $
(i.e. upon the conicity angle $\theta _{0}$ in the case of scattering
problems)\cite{cent1,ecossais}. Numerical and theoretical experiments have
shown that the dependence of the resonant wavelengths with respect of the
conicity angle is given by: 
\begin{equation}
\lambda _{r}(\theta _{0})=\lambda _{r}(90%
%TCIMACRO{\UNICODE[m]{0xb0}}%
%BeginExpansion
{{}^\circ}%
%EndExpansion
)\sin \theta _{0}  \label{reson}
\end{equation}
This behavior can be interpreted thanks to a simple model called ''model of
the infinitely conducting cavity'' \cite{cent2}. Although the localization
of light inside PCs doped by microcavities is a global phenomenon due to the
photonic band gap effect and the broken symmetry of the lattice, both
resonant wavelengths and localized modes may be well approximated using a
local representation of the electromagnetic field. Figure 5 presents the
diagram of transmission in the case of the doped PC without boundary for $3$
distinct conicity angles $90%
%TCIMACRO{\UNICODE[m]{0xb0}}%
%BeginExpansion
{{}^\circ}%
%EndExpansion
$, $80%
%TCIMACRO{\UNICODE[m]{0xb0}}%
%BeginExpansion
{{}^\circ}%
%EndExpansion
$, $70%
%TCIMACRO{\UNICODE[m]{0xb0}}%
%BeginExpansion
{{}^\circ}%
%EndExpansion
$ and for a fixed polarization angle $\delta _{0}=0%
%TCIMACRO{\UNICODE[m]{0xb0}}%
%BeginExpansion
{{}^\circ}%
%EndExpansion
$. The resonant wavelength is shifted toward the shorter wavelengths inside
the band gap when the conicity angle decreases. In table 1, both numerical
computations and predicted resonant wavelengths versus the conicity angle
are compared. The domain of validity of the ''infinitely conducting cavity''
model depends on the strength of the localization of the light with respect
to the conicity angle. For example in our case, decreasing the conicity
angle induced the shift of the localized modes toward the lower edge of the
band gap. Therefore the localized mode is delocalized out of the microcavity
hence the model of ''the infinitely conducting cavity'' is no more valid.
This behavior explains why equation (\ref{reson}) must be applied for low
inclinations of the wave vector with respect to the plane defined by the
cross section of the fibers (i.e. for $\theta <60%
%TCIMACRO{\UNICODE[m]{0xb0}}%
%BeginExpansion
{{}^\circ}%
%EndExpansion
$). Let us now study the optical properties of the same PC embedded inside
the cylinder sheath defined in fig.3 versus the conicity angle. Figure 8
presents the diagram of transmission obtained for $3$ values of the conicity
angle ($90%
%TCIMACRO{\UNICODE[m]{0xb0}}%
%BeginExpansion
{{}^\circ}%
%EndExpansion
$, $80%
%TCIMACRO{\UNICODE[m]{0xb0}}%
%BeginExpansion
{{}^\circ}%
%EndExpansion
$, $70%
%TCIMACRO{\UNICODE[m]{0xb0}}%
%BeginExpansion
{{}^\circ}%
%EndExpansion
$) and for a fixed polarization angle $\delta _{0}=0%
%TCIMACRO{\UNICODE[m]{0xb0}}%
%BeginExpansion
{{}^\circ}%
%EndExpansion
$. These results lead to two remarks:\newline
1) When decreasing the conicity angle, a second resonant wavelength appears,%
\newline
2) The shift of the resonant wavelength is weaker for a PC fiber than for a
PC without boundary.\newline
The apparition of the second resonant wavelength $\lambda _{r}^{c}(70%
%TCIMACRO{\UNICODE[m]{0xb0}}%
%BeginExpansion
{{}^\circ}%
%EndExpansion
)/R=2.226$ demonstrates that the optical interactions of the PC and the
cylinder sheath modify strongly the photonic band structures. Moreover, the
conicity angle dependence of the resonant wavelengths diminishes when the PC
is embedded inside a cylinder sheath. In that case, the model of the
''infinitely conducting cavity'' cannot be used to compute the shift of the
resonant wavelengths for grazing incidence. This behavior may be efficiently
used for controlling the detunning of the localized modes with respect to
the inclination of the wave vector.

\section{Conclusion}

We have extended the vector multi-scattering theory of diffraction by
parallel cylinders to the case of a hard boundary. We have applied this
theory in order to study the scattering properties of a PC embedded inside a
cylinder sheath. The numerical results have shown that PC fibers present
complicated photonic band structures with additional localized modes for
grazing incidence. The exterior circular boundary also attenuates the
detunning of the resonant wavelengths with respect to the inclination of the
wave vector (i.e. with the propagation coefficient $\gamma $). Matters
related to the optical effects of the size of the PC inside the cylinder
sheath will be discussed in future studies. This theory allows to study the
transmission properties of 2D PC with inverted contrast but it can also be
straightforwardly applied to the study of propagation phenomena in photonic
crystal fibers, where the modes are linked to the resonances of the
scattering matrix\cite{moi2,cent2,review} Work is also in progress in that
direction

{\bf Figures captions:}

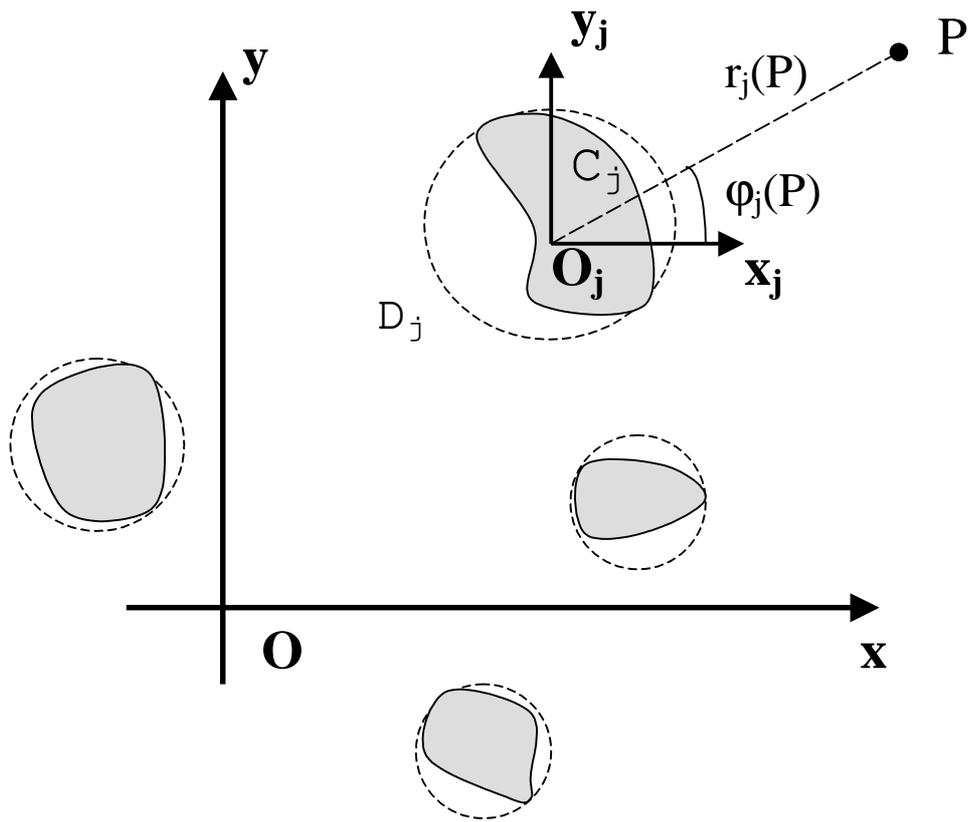
{\bf Figure 1:} Scattering by a set of parallel fibers of arbitrary shape,
optical index and position.

{\bf Figure 2}: Definition of the incident angle $\varphi _{0}$,
polarization angle $\delta _{0}$, conicity angle $\theta _{0}$ in the
Cartesian system $(O,x,y,z)$.

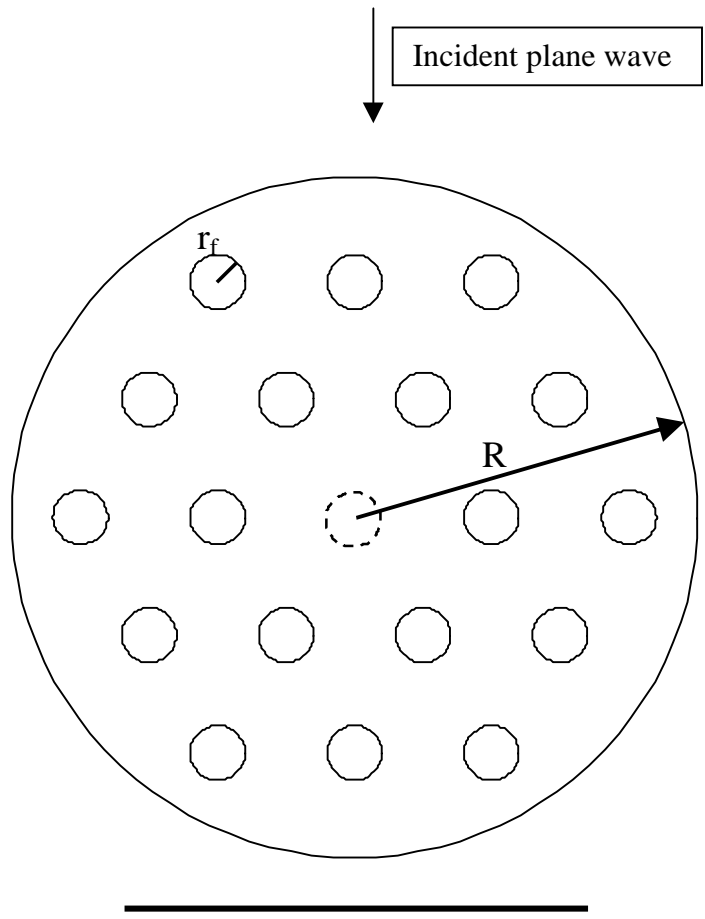
{\bf Figure 3: }Hexagonal PC constituted by $19$ air hole embedded in a
dielectric circular cylinder of optical index $\sqrt{\varepsilon _{r}}=4$.
When central air hole (dashed fiber) is removed the PC is doped by a
microcavity. The radii of the air hole and the cylinder sheath are
respectively $r_{f}=0.8$ and $R=10$. The segment below the structure is used
for the computation of the transmission coefficient $T$.

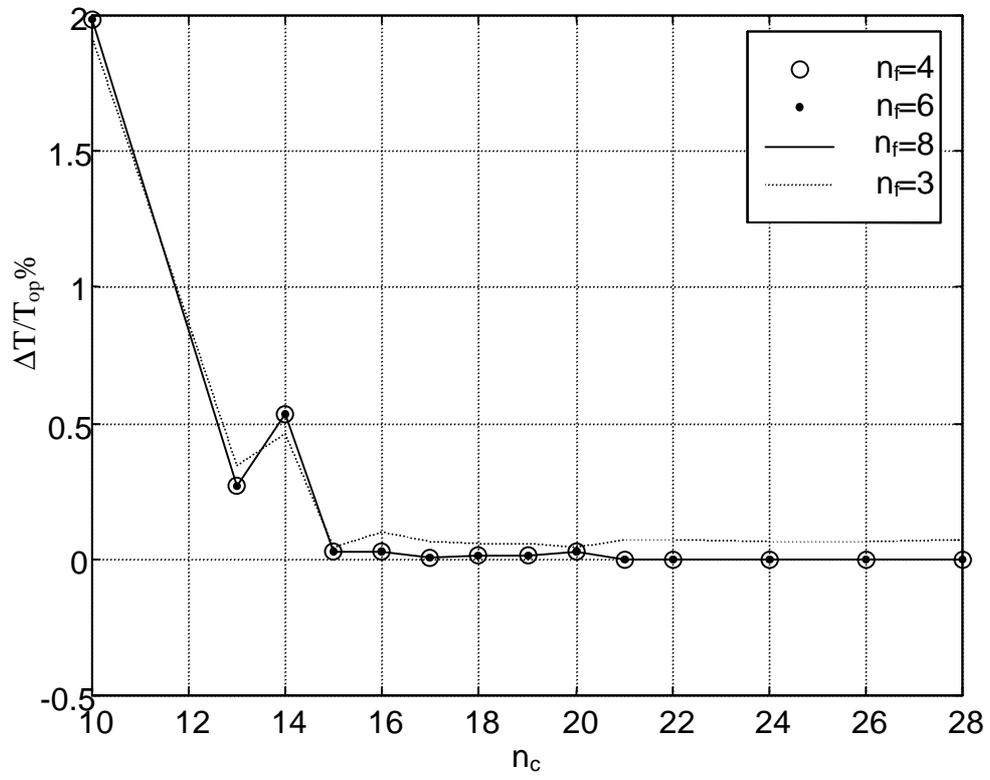
{\bf Figure 4}: Relative error of the transmission coefficient $T$ versus $%
n_{f}$ and $n_{c}$. The structure is illuminated by a plane wave in $p$%
-polarization and for $\lambda /R=1.70$.

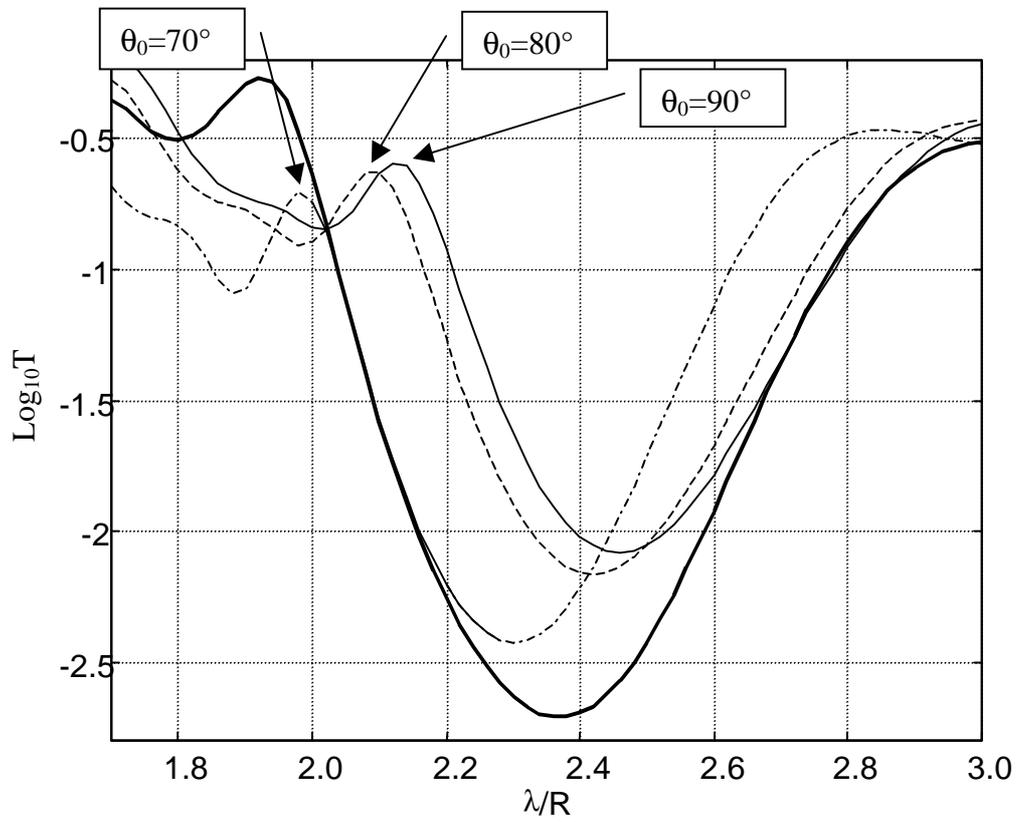
{\bf Figure 5}: Diagram of transmission for the PC defined in fig.3 but
without the circular cylinder sheath. The hexagonal PC\ presents a photonic
band gap approximately equal to $[2.15;2.70]$ for $p$-polarization (bold
curve). The solid, dashed and dotted-dashed curves are respectively obtained
for the incident field parameters: $\theta _{0}=[90%
%TCIMACRO{\UNICODE[m]{0xb0}}%
%BeginExpansion
{{}^\circ}%
%EndExpansion
;80%
%TCIMACRO{\UNICODE[m]{0xb0}}%
%BeginExpansion
{{}^\circ}%
%EndExpansion
;70%
%TCIMACRO{\UNICODE[m]{0xb0}}%
%BeginExpansion
{{}^\circ}%
%EndExpansion
]$ and for the same polarization angle $\delta _{0}=0%
%TCIMACRO{\UNICODE[m]{0xb0}}%
%BeginExpansion
{{}^\circ}%
%EndExpansion
$.

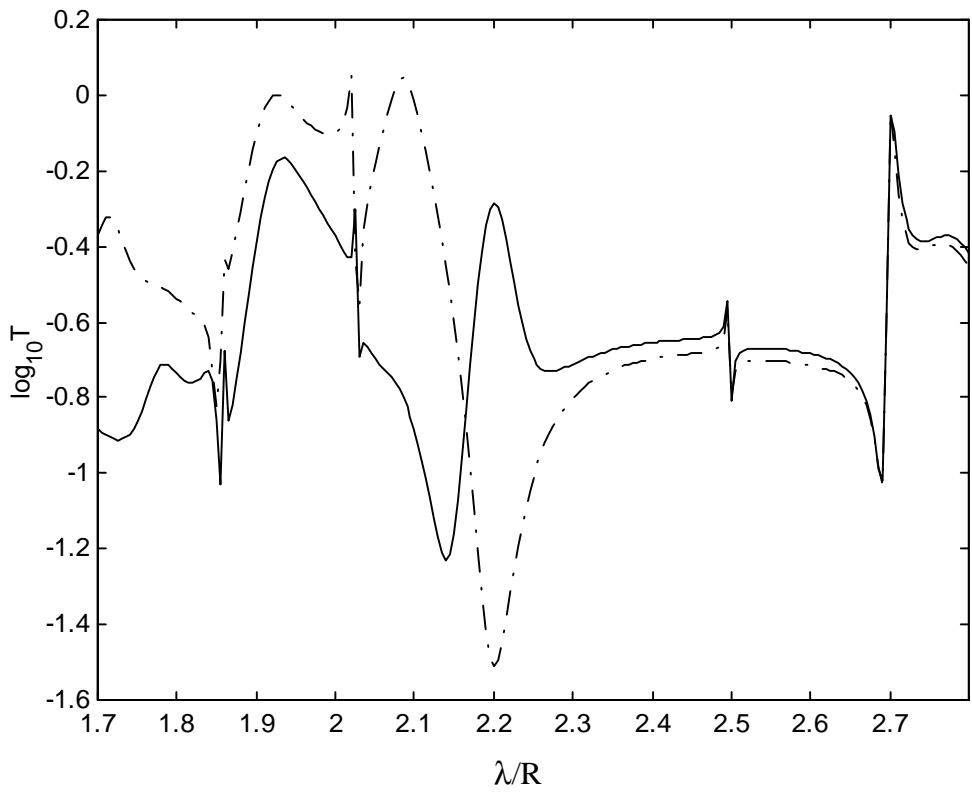
{\bf Figure 6}: Logarithm of transmission versus the wavelength for the
structure defined in fig.3 and for $p$-polarization: the solid curve is
obtained when the PC is doped by a central microcavity whereas the dashed
curve is computed for the perfect PC.

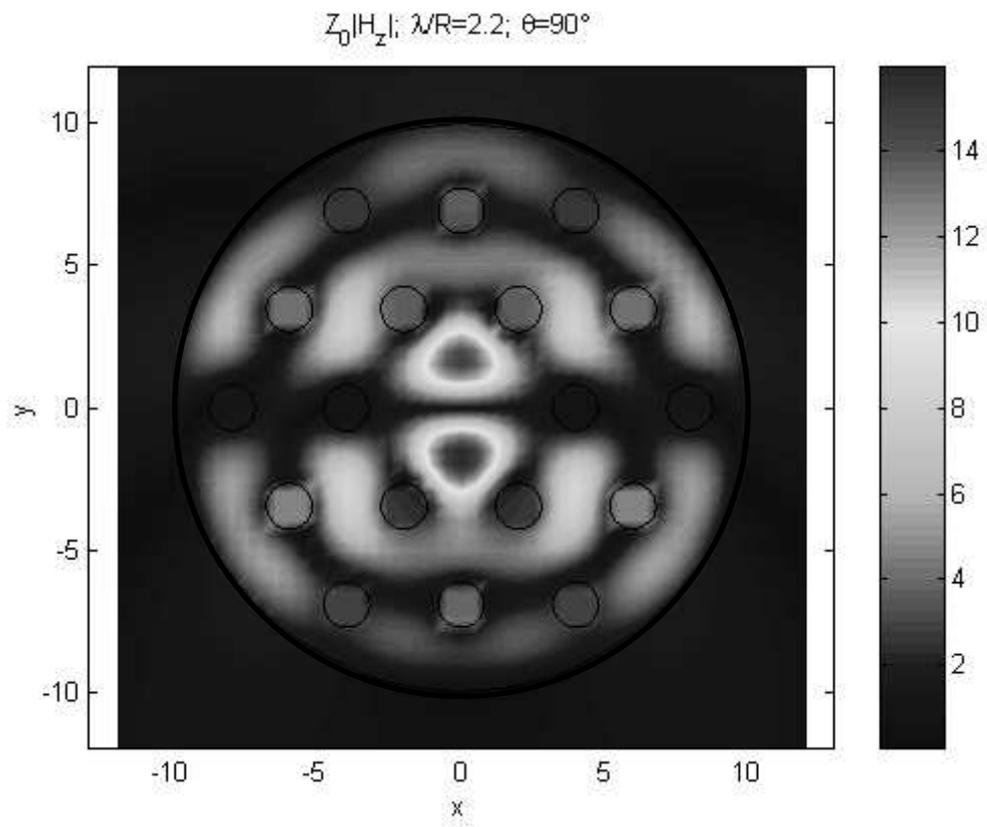
{\bf Figure 7}: Map of the modulus of the magnetic field for the wavelength
associated to the localized mode $\lambda _{r}^{c}(90%
%TCIMACRO{\UNICODE[m]{0xb0}}%
%BeginExpansion
{{}^\circ}%
%EndExpansion
)/R=2.20$.

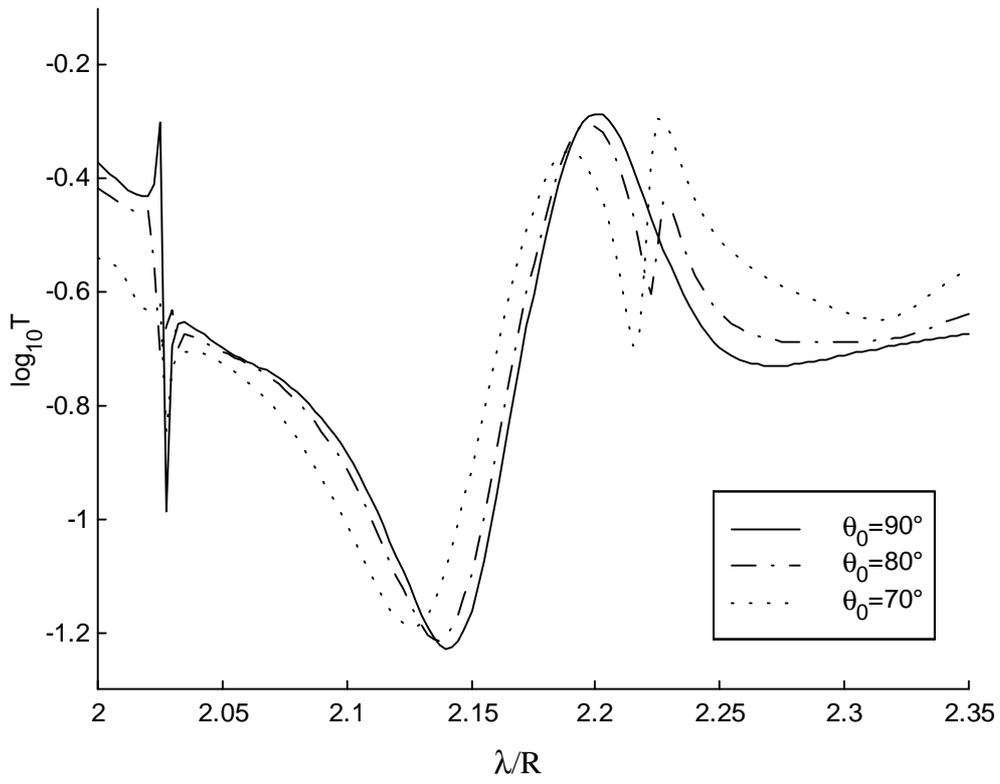
{\bf Figure 8}: Diagram of transmission for the structure of fig.3 for $3$
conicity angles $\theta _{0}=[90%
%TCIMACRO{\UNICODE[m]{0xb0}}%
%BeginExpansion
{{}^\circ}%
%EndExpansion
;80%
%TCIMACRO{\UNICODE[m]{0xb0}}%
%BeginExpansion
{{}^\circ}%
%EndExpansion
;70%
%TCIMACRO{\UNICODE[m]{0xb0}}%
%BeginExpansion
{{}^\circ}%
%EndExpansion
]$ and for a fixed polarization angle $\delta _{0}=0%
%TCIMACRO{\UNICODE[m]{0xb0}}%
%BeginExpansion
{{}^\circ}%
%EndExpansion
$.

{\bf Table caption:}

{\bf Table 1:} Comparison between the resonant wavelength versus the
conicity angle calculated with a direct numerical computation and thanks
equation (\ref{reson}).

\end{document}